\documentclass[aps,pre,longbibliography,notitlepage,floatfix,nofootinbib]{revtex4-1}

\usepackage{bm,amsmath,amssymb,graphicx,stmaryrd,float,subcaption,upgreek,sidecap,dsfont}
\usepackage{color}
\usepackage[abs]{overpic}
\usepackage{verbatim}
\usepackage{comment,footnote}
\usepackage[format=plain,labelfont={bf,small},textfont=small,justification=raggedright,singlelinecheck=false]{caption}

\newcommand{\bomega}{{\bm \omega}}
\newcommand{\bOmega}{{\bm \Omega}}
\newcommand{\bx}{{\bm x}}

\newcommand{\config}{{Y}}

\newcommand{\bn}{{\bm n}}
\newcommand{\bmo}{{\bm m}}
\newcommand{\bP}{{\bf P}}

\newcommand{\bS}{{\bf S}}

\newcommand{\bM}{{\bf M}}
\newcommand{\bv}{{\bm v}}

\newcommand{\bA}{{\bf A}}

\newcommand{\bB}{{\bf B}}

\newcommand{\bR}{{\bf R}}

\newcommand{\bQ}{{\bf Q}}

\newcommand{\jump}[1]{\llbracket#1\rrbracket}

\newcommand{\add}[1]{\langle#1\rangle}
\newcommand{\addb}[1]{\bigg\langle#1\bigg\rangle}
\newcommand{\addl}[1]{\langle#1\rangle_\lambda}
\newcommand{\addL}[1]{\langle#1\rangle_{1-\lambda}}

\newcommand{\zero}{{\bm 0}}

\begin{document}
	\title{Partial constraint singularities in elastic rods}
	\author{J. A. Hanna}
	\email{hannaj@vt.edu}
	\affiliation{Department of Biomedical Engineering and Mechanics, Department of Physics, Center for Soft Matter and Biological Physics, Virginia Polytechnic Institute and State University, Blacksburg, VA 24061, U.S.A.}
	\author{H. Singh}
	\email{harmeet@vt.edu}
	\affiliation{Department of Biomedical Engineering and Mechanics, Virginia Polytechnic Institute and State University, Blacksburg, VA 24061, U.S.A.}
	\author{E. G. Virga}
	\email{eg.virga@unipv.it}
	\affiliation{Dipartimento di Matematica, Universit\`a di Pavia, Via Ferrata 5, I-27100 Pavia, Italy}
	\date{\today}

\begin{abstract}
We present a unified classical treatment of partially constrained elastic rods. Partial constraints often entail singularities in both shapes and reactions. Our approach encompasses both sleeve and adhesion problems, and provides simple and unambiguous derivations of counterintuitive results in the literature.
Relationships between reaction forces and moments, geometry, and adhesion energies follow from the balance of energy during quasistatic motion.
We also relate our approach to the 
balance of material momentum and the concept of a driving traction. The theory is generalizable and can be applied to a wide array of contact, adhesion, gripping, and locomotion problems. 
\end{abstract}

\maketitle

\noindent Keywords: Rods, Constraints, Adhesion, Jump conditions \\
Mathematics Subject Classification: 74A15 (Thermodynamics), 74G70 (Stress concentrations, singularities), 74K10 (Rods), 74M15 (Contact)

\section{Introduction}\label{Sec:Introduction}

Partially constrained flexible structures provide fascinating everyday examples of the consequences of balance laws at geometric discontinuities, and their associated mechanical singularities.
Over the years, both theory and experiment have revealed challenging and counterintuitive results linking reaction forces and moments, adhesion energy, and curvatures of rods and strips at points of discontinuous contact with sleeves and sticky surfaces \cite{Bigoni15, Bottega90, Kinloch1994, Majidi2007, MajidiAdams09}. 
These results pertain to structural stability questions \cite{HumerIrschik11, Phungpaingam12}, paper handling and processing \cite{MansfieldSimmons87, StolteBenson93}, animal locomotion \cite{Majidi12, DeSimone15, CicconofriDeSimone15, DalCorso17}, standard tests for adhesives \cite{ASTMlooptack, Plaut2001, Plaut2001-2}, adhesion of lipid membranes and molecules \cite{Deserno07, GlassmakerHui04} or MEMS \cite{deBoerMichalske99, PampAdams07}, and
elastocapillarity \cite{Py09, HureAudoly2013, Elettro17}.

In this note, we present a simple classical derivation based on the local balance of force, moment, and energy that provides a unified framework for guide, contact, and adhesion problems for rods.  We apply this framework to two recent examples from the literature, involving sliding in or out of a frictionless sleeve \cite{Bigoni15} and peeling from an adhesive surface \cite{Majidi2007}.  We directly recover results on global force balances, and on boundary conditions relating reaction forces and moments, and bending and adhesion energies, avoiding assumptions and ambiguities present in prior derivations.
For simplicity of exposition, we focus our comments on an inextensible, unshearable, planar Euler \emph{elastica}, quasistatic dynamics, and discontinuities of second or higher order in position.  For inextensible, unshearable rods, this means discontinuities in curvature, but not kinks in tangent vectors; the bulk equivalent is an ``acceleration wave''.  However, there is nothing to prevent generalization of the present approach.

Other authors have taken several different approaches to partial constraint problems.  
Bigoni and co-workers \cite{Bigoni15} derived and experimentally confirmed the relationship between end loads on an \emph{elastica} partially contained, and free to slide within, a frictionless sleeve.  In their derivation, they employed an angle-pendulum representation of the shape equation, and extremized the elastic energy with respect to virtual shifts in the position of the sleeve edge.   One disadvantage of the commonly employed angle representation, in contrast to either a curvature representation or a rod mechanics approach featuring contact force and moment, is that it throws away local information about tangentially conserved quantities \cite{SinghHanna2017elastica}.  These quantities, the force and material force, both have prominent roles in rod problems and can be used to derive results more directly.  Bigoni and co-workers also obtained a relation between the reaction force and moment at the sleeve edge using an argument that considers a limiting geometry of the sleeve.  
They interpret part of the reaction force as an ``Eshelby-like'' term analogous to a configurational force.  
However, O'Reilly \cite{OReilly2015Eshelby} has shown that the results follow from an \emph{absence} of a reaction ``material force'', a configurational source term arising in his material momentum balance approach.
His derivation is elegant and direct, but non-classical in that it requires prescription of a singular source of material force, or lack thereof, at the contact boundary.  
In contrast with the familiar concept of an energy source or sink, our understanding of sources of material force, and how to prescribe such a source for a given physical situation, is still being developed.  
For systems governed by an action principle, material forces in the bulk can be understood as conserved quantities associated with symmetries in material properties of the rod; for purely mechanical systems such as an \emph{elastica} without adhesion, the material momentum balance is simply the projection of the linear momentum balance onto the tangents of the body \cite{SinghHanna2017elastica}.
The literature suggests that discontinuities in material properties or internal energy can serve as sources of material force \cite{KienzlerHermann1986,KienzlerHerrmann00,Maugin11, OReilly2017}.  It seems clear that for the problem of a uniform rod in a frictionless sleeve, no material source term is present at the contact discontinuity, but we will see later that it is not obvious how to prescribe this term for problems involving adhesion.
Majidi and co-workers examined adhesion of an \emph{elastica} from a variety of perspectives, including that of O'Reilly's material momentum balance, for which they interpreted the material source term as the jump in adhesion energy across the peeling front \cite{Majidi2007, Majidi12, Majidi13}.  This approach led them to a boundary condition with an extra term, which they then assumed to vanish in order to agree with their own alternate derivations and with established results relating the reaction moment and bending energy at the peeling front.

Here we will show that all of the above results can be derived from classical balance laws at a moving geometric discontinuity. 
These jump conditions are introduced, along with our notation and other preliminaries, in Section \ref{Sec:balancelaws}. They are also derived in Section \ref{Sec:invariance} from invariance considerations applied to the internal energy balance at a discontinuity. 
We there also discuss the free energy imbalance at a discontinuity for adiabatic processes, 
derive an expression for dissipation at the singular point and relate it to the concept of a driving traction, and obtain constitutive restrictions on free energy 
from the second law of thermodynamics.
In Section \ref{Sec:solutions}, we identify the conserved force and material force, and solve sleeve and adhesion problems for an \emph{elastica}. 
In particular, we show that the natural boundary conditions of interest follow from the balance of energy during a quasistatic motion.
We conclude with a comparison of our approach and the use of a material momentum balance, and some additional comments, in Sections \ref{Sec:material_force} and \ref{Sec:discussion}.  While for simple mechanical systems the two approaches provide essentially the same information, our energy balance approach unambiguously avoids the appearance of a phantom tangential reaction term in the adhesion problem.

\section{Quasistatic balance laws}\label{Sec:balancelaws}

We first present the kinematics and bulk and singular balance laws for an inextensible rod in quasistatic equilibrium, neglecting inertial effects.

Consider a planar, twistless rod described as a curve $\bx(s,t)$ parameterized by a material coordinate $s\in [0,L]$ and time $t$, with a single discontinuity in field quantities at the time-dependent non-material point $s=s_0(t)$.
A prime $(\,)'$ will denote an $s$-derivative, and a dot $\dot{(\,)}$ a material time derivative when applied to quantities depending on $s$ and a partial time derivative when applied to a non-material quantity such as $s_0(t)$.  Thus, $\dot{s}_0$ denotes how fast the discontinuity moves through the material composing the rod. 
For an inextensible, unshearable rod, $s$ is also the arc length and $\bx'$ a unit tangent vector.  For compactness of expressions, we will use the shorthand $\bomega\equiv\bx'\times\dot{\bx}'$ for angular velocity and $\bOmega\equiv\bx'\times\bx''$ for the Darboux vector.
The jump and mean of a quantity $\bQ$ across the discontinuity $s_0$ are denoted by $\jump{\bQ}\equiv \bQ^+ - \bQ^-$ and $\add{\bQ}\equiv\tfrac{1}{2}(\bQ^+ + \bQ^-)$, where $\bQ^\pm$ are the values of $\bQ$ immediately on either side of the discontinuity.

In the absence of body or applied forces or couples, the force and moment balances in the bulk of the rod are given in terms of a contact force $\bn$ and contact moment $\bmo$ by \cite{AntmanElasticity1995}
\begin{align}
\bn'&=\bf 0\, ,\label{force_balance_bulk}\\
\bmo'+\bx'\times\bn&=\bf 0\, .\label{moment_balance_bulk}
\end{align}
 We will refer to four jump conditions at the point of discontinuity $s=s_0(t)$,
\begin{align}
\bR + \jump{\bn}&=\bf 0\, ,\label{force_jump}\\
\bM+\jump{\bmo}&=\bf 0\, ,\label{moment_jump}\\
\tilde{E} + \jump{\bn\cdot\dot{\bx} + \bmo\cdot\bomega+ \dot{s}_0\varepsilon}&=0\, ,\label{energy_jump}\\
\config + \jump{ \bn\cdot\bx' + \bmo\cdot\bOmega - \psi} &= 0\, ,\label{material_force_jump}
\end{align}
where $\varepsilon$ and $\psi$ are the internal energy and the free energy, respectively, per unit 
length of the rod.
The first three conditions (\ref{force_jump})-(\ref{energy_jump}) are classical force, moment, and energy jump conditions, with allowance for singular sources of force $\bR$, moment $\bM$, and power $\tilde{E}$ \cite{GreenNaghdi78, OReillyVaradi99, OReilly07, Hanna15}.  We will use these conditions to solve  the specific problems to which we apply our theory in Section \ref{Sec:solutions}. Note that $\tilde{E}$ is the  power input at the singularity; it should not be confused with the \emph{dissipation} $\mathcal{D}$ at the singularity, which will be introduced in Section \ref{Sec:invariance} below.  
The fourth, non-classical, condition \eqref{material_force_jump} is a material momentum jump condition with a singular source of material force $\config$ \cite{OReilly2007, OReilly2017}.  We will not use this condition, but will discuss its significance in Section \ref{Sec:material_force}.  In fact, the bracketed term in \eqref{material_force_jump} can be given a classical interpretation for many simple conservative systems, in terms of material symmetry of an action \cite{SinghHanna2017elastica} or the quasistatic form of a conservation law for uniform hyperelastic rods \cite{MaddocksDichman1994}, but we lack an unambiguous way to specify the source $\config$.  
Fully dynamic forms of the jump conditions (\ref{force_jump})-(\ref{material_force_jump}) can be found elsewhere \cite{OReilly07, Hanna15}.  
Boundary conditions on force and moment simply involve the bracketed $\bn$ and $\bmo$ terms in (\ref{force_jump})-(\ref{moment_jump}).  For example, if an end load $\bP$ is applied at $s=L$, with the body lying on the $\,^-$ side, we would write $\bP - \bn^-(L)=\zero$. Similarly, with zero applied end moment, we would write $-\bmo^-(L)=\zero$.
 
We assume continuity of the position vector, $\jump{\bx\left(s_0(t),t\right)}=\zero$, and the tangents, $\jump{\bx'\left(s_0(t),t\right)}=\zero$, across the point of discontinuity.  Total time derivatives of these expressions, and subsequent application of $\bx' \times$ to the second, give the following kinematic jump conditions,
\begin{align}
\jump{\dot{\bx} + \dot{s}_0\bx'}&={\bf 0}\, ,\label{kinematic_compatibility}\\
\jump{\dot{\bx}' + \dot{s}_0\bx''}&= \bf 0\, ,\label{kinematic_compatibility_2}\\
\jump{\bomega+ \dot{s}_0\bOmega}&=\bf 0\, .\label{angular_compatibility}
\end{align}
The first and third of these express continuity of linear and angular velocity of the non-material point of discontinuity,
\begin{align}
\bv_0 &= \dot{\bx}^\pm + \dot{s}_0\dot{\bx}'^\pm \,\, = \add{\dot{\bx}} + \dot{s}_0\add{\bx'} =\dot{\bx}(s_0) + \dot{s}_0 \bx'(s_0)  \, ,\label{linear_velocity_discontinuity}\\
\bomega_0 &= \bomega^\pm + \dot{s}_0\bOmega^\pm = \add{\bomega}+\dot{s}_0\add{\bOmega}\, .\label{angular_velocity_discontinuity}
\end{align}

\section{Dissipation and invariance}\label{Sec:invariance}
Here, following the pattern set forth by Noll \cite{noll:mecanique}, we prove that the singular balance laws \eqref{force_jump}-\eqref{energy_jump} can all be derived from an invariance property required of the balance of energy at the singularity. We also consider the free energy imbalance to arrive at restrictions on the constitutive assumptions made for the rod.

\subsection{Energy balance}\label{Sec:energy_balance}

For a rod with an energy source only at the singularity, and no heat source, we may write the energy balance as
\begin{align}
\frac{d}{dt}\int_{s_1}^{s_2}\!\! ds\, \varepsilon = E + W + \left.\bn\cdot\dot{\bx}\,\right|_{s_1}^{s_2} + \left.\bmo\cdot\bomega\,\right|_{s_1}^{s_2} \, , \label{C-D1}
\end{align}
for a material interval enclosing a moving discontinuity at $s_0(t)$, such that $s_0(t)\in[s_1, s_2]$. In the expression above, $E$ represents a source of power at the singularity, and
\begin{equation}\label{eq:W}
W=\bR\cdot\add{\dot{\bx}} + \bM\cdot\add{\bomega}
\end{equation}
is the power expended there by the singular sources of force and moment.

In principle, $W$ could instead be prescribed 
 in a more general form,
\begin{equation}\label{eq:W_general}
W=\bR\cdot\addl{\dot{\bx}} + \bM\cdot\addl{\bomega},
\end{equation}
where, for any $\lambda\in[0,1]$, we define the convex combination $\addl{\,}=\lambda(\,)^++(1-\lambda)(\,)^-$. Clearly, $\add{\,}$ corresponds to $\addl{\,}$ for $\lambda=\frac12$, and it is the only kinematic measure $\addl{\,}$ symmetric under the exchange of end signs $\pm$. For the sake of generality, and only in this section, we shall write $W$ as in \eqref{eq:W_general}.

Splitting the integral in \eqref{C-D1} over two time-dependent intervals, applying the Leibniz rule, and letting $s_1$ and $s_2$ approach $s_0$ from below and above, respectively, we obtain
\begin{align}\label{eq:energy_singular_balance}
E+\bR\cdot\addl{\dot{\bx}} + \bM\cdot\addl{\bomega}+\jump{\bn\cdot\dot{\bx} + \bmo\cdot\bomega + \dot{s}_0\varepsilon} =0\, ,
\end{align}
which coincides with \eqref{energy_jump}, provided that we set
\begin{equation}
\tilde{E}= E  + \bR\cdot\addl{\dot{\bx}} + \bM\cdot\addl{\bomega}\, .\label{E_definition_1}
\end{equation}
Equation \eqref{E_definition_1} expresses the balance of working at the singularity; it reveals $\tilde{E}$ as the \emph{net} singular power, comprising a source $E$ and the working of the singular sources $\bR$ and $\bM$.

We now require that the singular balance \eqref{eq:energy_singular_balance} be invariant under translations and rotations, assuming that both $E$ and $\varepsilon$  represent objective quantities. Transforming $\dot{\bx}\mapsto\dot{\bx}+\bm{v}^\ast$, $\bomega\mapsto\bomega+\bm{\omega}^\ast$, 
with arbitrary vectors $\bm{v}^\ast$ and $\bomega^\ast$, the former in the plane of the rod and the latter orthogonal to the plane, we easily see that \eqref{eq:energy_singular_balance} is also valid for the transformed motion, provided that
\begin{equation}\label{eq:compatibility_condition}
\big(\jump{\bn}+\bR\big)\cdot\bm{v}^\ast+\big(\jump{\bm{m}}+\bM\big)\cdot\bomega^\ast=0\, ,
\end{equation}
from which, by the arbitrariness of $\bm{v}^\ast$ and $\bomega^\ast$, both \eqref{force_jump} and \eqref{moment_jump} follow at once.\footnote{This argument is presented for a planar problem, but can be fully generalized.} We now use these equations to arrive at a reduced form of the singular balance of energy \eqref{eq:energy_singular_balance}. Using 
the algebraic identity
\begin{equation}\label{eq:algebraic_identity}
\jump{\bA\cdot\bB}=\jump{\bA}\cdot\addl{\bB} + \addL{\bA}\cdot\jump{\bB}\, ,
\end{equation}
equation \eqref{eq:energy_singular_balance} is given the form
\begin{align}
E  = -\addL{\bn}\cdot\jump{\dot{\bx}} - \addL{\bmo}\cdot\jump{\bomega}- \dot{s}_0\jump{\varepsilon}\, .\label{energy_jump_11}
\end{align}
Using the linear and angular kinematic jump conditions \eqref{kinematic_compatibility} and  \eqref{angular_compatibility}  to substitute for $\jump{\dot{\bx}}$ and $\jump{\bomega}$, we  rearrange the latter equation as  
\begin{align}
E  = \dot{s}_0\big(\addL{\bn}\cdot\jump{\bx'} + \addL{\bmo}\cdot\jump{\bOmega} - \jump{\varepsilon}\big)\, ,\label{driving_traction_1}
\end{align}
which is our \emph{reduced} singular energy balance.

Although these conclusions were achieved for the general definition of $W$ in \eqref{eq:W_general}, in the following  we shall take the familiar value $\lambda=\frac12$, but only for simplicity. Thus, equations \eqref{driving_traction_1} and \eqref{E_definition_1} become
\begin{align}
E  = \dot{s}_0\big(\add{\bn}\cdot\jump{\bx'} + \add{\bmo}\cdot\jump{\bOmega} - \jump{\varepsilon}\big)\label{driving_traction}
\end{align}
and
\begin{equation}\label{eq:A1}
\tilde{E}= E  + \bR\cdot\add{\dot{\bx}} + \bM\cdot\add{\bomega}\, , 
\end{equation}
respectively. Both of these equations are consequences of \eqref{C-D1}, which henceforth  will be assumed to be valid.

Note that \eqref{driving_traction} could also be obtained by splitting the kinematic quantities in \eqref{energy_jump} into invariant and non-invariant parts. Recall that with our continuity assumptions, the linear and angular velocities \eqref{linear_velocity_discontinuity} and \eqref{angular_velocity_discontinuity} of the non-material point have the forms $\add{\dot{\bx}} + \dot{s}_0\add{\bx'}$ and $\add{\bomega} + \dot{s}_0\add{\bOmega}$. Using the identity \eqref{eq:algebraic_identity} with $\lambda=\frac{1}{2}$, \eqref{energy_jump} can be written as $\tilde{E} + \jump{\bn}\cdot\add{\dot{\bx}}+\jump{\bmo}\cdot\add{\bomega} = -\add{\bn}\cdot\jump{\dot{\bx}} - \add{\bmo}\cdot\jump{\bomega} - \dot{s}_0\jump{\varepsilon}$, which upon using the kinematic conditions  \eqref{kinematic_compatibility} and \eqref{angular_compatibility} and the force and moment conditions \eqref{force_jump} and \eqref{moment_jump} reduces to
\begin{align}
\tilde{E}-\bR\cdot\add{\dot{\bx}}-\bM\cdot\add{\bomega} = \dot{s}_0\big(\add{\bn}\cdot\jump{\bx'} + \add{\bmo}\cdot\jump{\bOmega} - \jump{\varepsilon}\big)\, ,\label{energy_balance_splitting}
\end{align}
which corresponds to \eqref{driving_traction} and \eqref{eq:A1}.
The splitting of the energy balance in \eqref{energy_balance_splitting} has paired non-invariant terms with the reaction force $\bR$ and moment $\bM$, and  invariant terms with the average contact force $\add{\bn}$ and moment $\add{\bmo}$.

A further assumption will be made in solving the  equilibrium problems detailed in the following section, namely that the net singular power input $\tilde{E}$ vanishes,
\begin{equation}\label{eq:extra_assumption}
\tilde{E}=0\,.
\end{equation}
Assumptions \eqref{C-D1} and \eqref{eq:extra_assumption} have a completely different status: the former is a general energy balance for rods with a (possibly dissipative) singularity, the latter, combined with \eqref{eq:A1}, represents a \emph{detailed} energy balance at the singularity, which may or may not be valid in general. The equilibrium problems we consider will also elucidate the different roles that the reduced singular energy balance \eqref{driving_traction} can play in our theory.  It may either be an identity expressing compatibility between the singular balances of force, moment, and energy, or a prescription for the energy source $E$.

\subsection{Free energy imbalance}\label{Sec:free_energy_imbalance}
We will use the free energy imbalance of a rod undergoing an adiabatic process to define the dissipation at the singularity.  In section \ref{Sec:adhesion}, the inferences drawn in the present section will help us gain insight into the physical meaning of the term $E$ in a more general thermodynamic setting.

As we proceeded for the energy balance in \ref{Sec:energy_balance}, 
we may write the free energy imbalance for a material interval enclosing a moving discontinuity as
\begin{align}
\frac{d}{dt}\int_{s_1}^{s_2}\!ds\,\psi  + \int_{s_1}^{s_2}\! ds\,\dot{\theta}\eta= -\mathcal{D} + W + \bn\cdot\dot{\bx}|_{s_1}^{s_2} + \bmo\cdot\bomega|_{s_1}^{s_2}\, ,\quad \mathcal{D}\ge 0 \, ,
\end{align}
where $\mathcal{D}$ is the dissipation at the singularity. The free energy $\psi$ is related to the internal energy $\varepsilon$, entropy $\eta$, and temperature $\theta$ through the relation $\psi = \varepsilon - \theta\eta$. Localizing this imbalance by the same procedure as before, we obtain
\begin{align}
\mathcal{D} = W + \jump{\bn\cdot\dot{\bx} + \bmo\cdot\bomega + \psi}\, ,\quad\mathcal{D}\ge 0\, ,\label{dissipation_jump}
\end{align}
which upon using \eqref{eq:W} and some algebraic manipulations can be written as
\begin{align}
\mathcal{D} = -\dot{s}_0(\add{\bn}\cdot\jump{\bx'} + \add{\bmo}\cdot\jump{\bOmega} - \jump{\psi})\, .\label{dissipation_jump_2}
\end{align}
The quantity appearing in \eqref{dissipation_jump_2} conjugate to the motion $\dot{s}_0$ of the singularity through the body is the driving traction of Abeyaratne and Knowles \cite{Abeyaratne1990, Abeyaratne1991}.

We prescribe the following expressions for the free energy and entropy,
\begin{align}
\psi &= \hat{\psi}(\bOmega,\theta) + \psi_c\, ,\label{free_energy_prescription_general}\\
\eta &= \hat{\eta}(\bOmega,\theta) + \eta_c\, .\label{entropy_prescription_general}
\end{align}
The internal energy can be determined as $\varepsilon=\hat{\varepsilon}(\bOmega,\theta) + \varepsilon_c$, where $\varepsilon_c = \psi_c + \theta\eta_c$. The $\psi_c$ and $\eta_c$ are constants whose jumps will account for jumps in the \emph{descriptions} of free energy and entropy across the singularity, as will be necessary when we discuss peeling in Section \ref{Sec:adhesion}.

Without loss of generality, we assume a moving discontinuity and parameterize the rod such that $-\dot{s}_0>0$.
Using the continuity of tangents $\jump{\bx'}=0$ and the definition of contact moment  for a hyperelastic rod $\bmo=\tfrac{\partial\psi}{\partial\bOmega} $, and substituting \eqref{free_energy_prescription_general} in \eqref{dissipation_jump_2} we obtain for any admissible quasistatic motion,
\begin{align}
\addb{\frac{\partial\hat{\psi}}{\partial\bOmega}}\cdot\jump{\bOmega} - \jump{\hat{\psi}} - \jump{\psi_c}\ge 0\, .\label{dissipation_jump_3}
\end{align}
Returning to the definition of $E$, we use $\jump{\varepsilon} = \jump{\hat{\psi}} + \jump{\varepsilon_c} + \jump{\theta\hat{\eta}}$ and the definition of the contact moment $\bmo$ to re-write \eqref{driving_traction} as
\begin{align}
E = \dot{s}_0 \bigg(\add{\bn}\cdot\jump{\bx'} + \addb{\frac{\partial \hat{\psi}}{\partial\bOmega}}\cdot\jump{\bOmega} - \jump{\hat{\psi}} - \jump{\varepsilon_c} - \jump{\theta\hat{\eta}}\bigg)\, .\label{E_general_1} 
\end{align}
The first term inside the round brackets vanishes due to continuity of tangents, while \eqref{dissipation_jump_3} simplifies the other terms such that we obtain the inequality
\begin{align}
E \le -\dot{s}_0\jump{\theta\eta}\, .\label{E_general_2}
\end{align}

\section{Equilibrium problems}\label{Sec:solutions}

We now derive results pertaining to sleeve and adhesion problems in quasistatic equilibrium.
We first identify two useful conserved quantities.  While, for the sleeve example, these quantities could be derived from spatial and material symmetries of an action \cite{SinghHanna2017elastica}, we will obtain them directly from the balances  (\ref{force_balance_bulk})-(\ref{moment_balance_bulk}).

Clearly, the contact force in \eqref{force_balance_bulk} is conserved,
\begin{align}
\bn = \bP\, ,\label{conservation_n}
\end{align}
in which expression we anticipate identifying the constant of integration with an end load $\bP$ at $s=L$. 
Another conserved quantity arises by integrating the tangential projection of \eqref{force_balance_bulk}.  We write $\bn'\cdot\bx' = (\bn\cdot\bx')' - \bn\cdot\bx''=0$ and project the moment balance \eqref{moment_balance_bulk} onto the (planar, twistless) Darboux vector $\bOmega = \bx'\times\bx''$ to obtain $\bn\cdot\bx'' = -\bmo'\cdot\bOmega$, making use of the orthogonality of $\bx'$ and $\bx''$.
Integrating, we obtain a constant $c$,
\begin{align}
\bn\cdot\bx' + \bmo\cdot\bOmega - \int\!\! ds\, \bmo\cdot\bOmega' = c \, . \label{conservation_C_step_4}
\end{align}
For a hyperelastic rod, the contact moment can be defined as $\bmo = \tfrac{\partial\psi}{\partial\bOmega}$, which reduces \eqref{conservation_C_step_4} to
\begin{align}
\bn\cdot\bx' + \bmo\cdot\bOmega - \psi = c \, .\label{C_general}
\end{align}
The conserved quantity $c$ is the ``material force"\footnote{O'Reilly  \cite{OReilly07,OReilly2015Eshelby} uses terms which can be easily translated into ours as $\mathsf{C} = -c$ and $\mathsf{B} = -\config$.}
 \cite{OReilly07,OReilly2015Eshelby}, which for a purely mechanical system corresponds to a symmetry with respect to constant shifts in the material coordinate $s$ \cite{Broer70, MaddocksDichman1994, SinghHanna2017elastica}.  
It is the same quantity that occurs in the material momentum jump condition \eqref{material_force_jump}.  In general, a corresponding bulk balance law for this quantity can also be derived, but for purely mechanical systems it simply reduces to the tangential projection of the conservation law for linear momentum.  For the present purposes, the interpretation of $c$ is irrelevant.

With the conserved quantities $\bP$ and $c$ in hand, the solution of partial contact problems is simple and direct, as will be shown presently.

\subsection{A sleeve problem}\label{Sec:sleeve}

\begin{figure}[h!]
	\includegraphics[width=12cm]{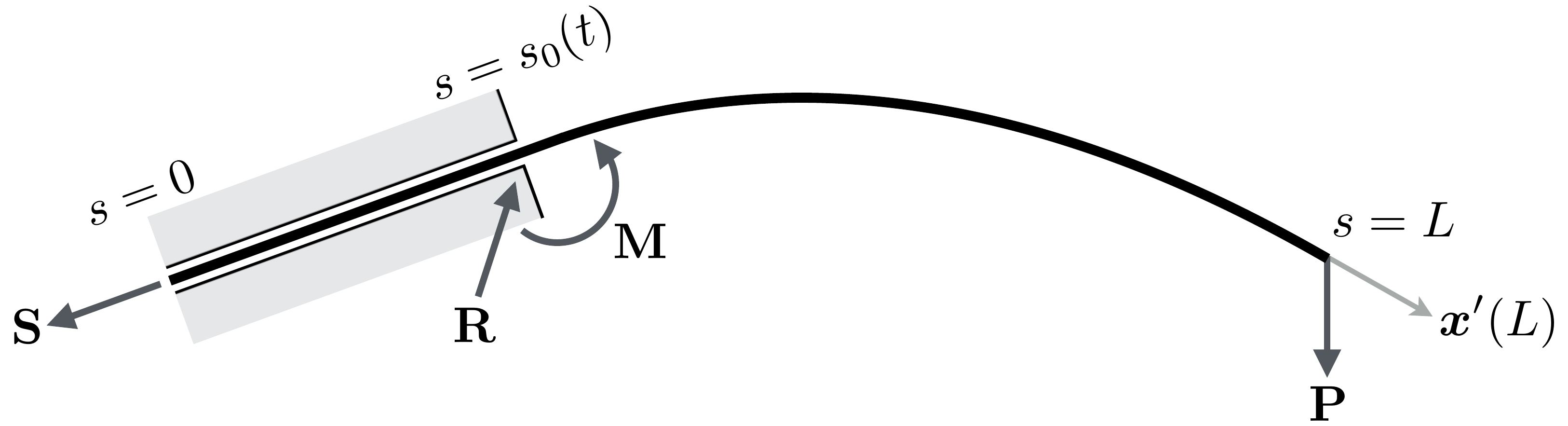}
	\captionsetup{margin=3cm}
	\caption{A rod $\bx(s,t)$ partially constrained by a frictionless sleeve and subject to end loads $\bS$ and $\bP$.  A reaction force $\bR$ and moment $\bM$ are present at the sleeve edge $s=s_0(t)$.}
	\label{Fig:elastica_setup}
\end{figure}

Consider the arrangement in Figure \ref{Fig:elastica_setup}, a problem studied by Bigoni and co-workers in \cite{Bigoni15}. 
A rod is bilaterally constrained by a straight, frictionless sleeve for $s\in[0,s_0]$, while two forces $\bS$ and $\bP$ are respectively applied to the ends at $s=0$ and $s=L$, with $\bS$ being parallel to the sleeve and rod. We will treat the system as purely mechanical by ignoring entropic terms, and assuming no dissipation at the sleeve edge, $\mathcal{D}=0$. With these assumptions, the free and internal energies are indistinguishable, and the results on this problem do not rely on the theory presented in Section \ref{Sec:free_energy_imbalance}.

The rod is an \emph{elastica} with uniform bending modulus $B$ and the constitutive prescription
\begin{align}
\psi = \hat{\psi} = \tfrac{1}{2}B\bx''\cdot\bx'' = \tfrac{\bmo\cdot\bmo}{2B}\, , \quad\bmo=B\bOmega\, ,\label{elastica_prescriptions}
\end{align}
so that condition \eqref{dissipation_jump_3} is identically satisfied, and expression \eqref{C_general} becomes
\begin{align}
\bn\cdot\bx' + \tfrac{\bmo\cdot\bmo}{2B} = c \, .\label{conservation_C}
\end{align}

As shown in \cite{Bigoni15}, the equilibrium relationship between the loads turns out to be $-\bS\cdot\bx'(0)=\bP\cdot\bx'(L)$.
From our perspective, we can immediately see that this looks like a statement about conservation of the material force or the tangential projection of linear momentum, but it's not apparent that this should hold in the presence of a reaction force $\bR$ and a reaction moment $\bM$ at the sleeve edge.  Indeed, one might be tempted to assume that a frictionless sleeve could only exert a reaction force $\bR$ without any tangential component, but it turns out that this is not the case, and the magnitude of $\bM$ and the tangential component of $\bR$ are intimately related.

The rather idealized assumptions we are using are that the constrained part of the rod is perfectly straight and the sleeve perfectly frictionless, with no projection of external forces on the rod tangent.  Thus, $\bmo^- =\zero$, and \eqref{conservation_C} tells us that $\bn^-\cdot\bx'(s_0) = -\bS\cdot\bx'(0)$.  We also note that any forces on the rod in the sleeve must be equal and opposite normal forces, and thus the force balance \eqref{force_balance_bulk} tells us that $\bn^- = -\bS$ and $\bn^+ = \bP$. 
The force and moment jump conditions \eqref{force_jump} and \eqref{moment_jump} at the sleeve edge can be written as
\begin{align}
\bR + \bP + \bS=\zero\, ,\label{force_balance_sleeve_edge}\\
\bM + \bmo^+ = \zero\, .\label{moment_balance_sleeve_edge}
\end{align}
Equation \eqref{force_balance_sleeve_edge} is also a global force balance. 
Using \eqref{conservation_C} on the free portion of the rod yields
 $\bP\cdot\bx'(L) + \tfrac{\bmo(L)\cdot\bmo(L)}{2B} = \bP\cdot\bx'(s_0)+\tfrac{\bmo^+\cdot\bmo^+}{2B}$, which in the current absence of an applied end moment is just
\begin{align}
\bP\cdot\bx'(L)  = \bP\cdot\bx'(s_0)+\tfrac{\bmo^+\cdot\bmo^+}{2B}\, .\label{projections_relation}
\end{align}
Projecting \eqref{force_balance_sleeve_edge} on the tangent $\bx'(s_0)$ at the sleeve edge, and using \eqref{projections_relation} and \eqref{moment_balance_sleeve_edge}, we obtain the relation
\begin{align}
-\bS\cdot\bx'(0)
 - \bP\cdot\bx'(L) = \bR\cdot\bx'(s_0) - \tfrac{\bM\cdot\bM}{2B}\, .
\label{force_balance_sleeve_edge_2}
\end{align}
This tangential force balance has been arranged so as to place the unknown reaction terms on the right hand side.  We now obtain further information about these terms by employing the energy balance.
By the continuity of tangents, and the constitutive prescriptions \eqref{elastica_prescriptions} for the \emph{elastica}, the right hand side of \eqref{driving_traction} vanishes, as it should since no additional mechanism to inject or absorb power is envisioned here. If, in addition, we assume that the \emph{net} power input across the sleeve edge vanishes, 
$\tilde{E}=0$, then the expression \eqref{eq:A1} reduces to
\begin{align}
W = \bR\cdot\add{\dot{\bx}} +\bM\cdot\add{\bomega}=0\, ,\label{energy_jump_2}
\end{align}
which has the transparent mechanical meaning of requiring the sleeve  \emph{edge} to exert a powerless system of reactions.
The non-material point representing the edge neither translates nor rotates
($\bv_0=\zero$ and $\bomega_0=\zero$), and the curvature inside the sleeve vanishes ($\bx''^-=\zero$), so the kinematics \eqref{linear_velocity_discontinuity} and  \eqref{angular_velocity_discontinuity} tell us that $\add{\dot{\bx}}=-\dot{s}_0\bx'(s_0)$ and $\add{\bomega}=-\dot{s}_0\tfrac{1}{2}\bOmega^+$.  These, in conjunction with the constitutive prescription for the moment \eqref{elastica_prescriptions} and the moment jump  \eqref{moment_balance_sleeve_edge}, transform \eqref{energy_jump_2} into
\begin{align}
\dot{s}_0\big(\bR\cdot\bx'(s_0) - \tfrac{\bM\cdot\bM}{2B}\big) = 0\, .\label{energy_balance_sleeve_edge_1}
\end{align}
Thus, for any quasistatic motion, 
\begin{align}
\bR\cdot\bx'(s_0) = \tfrac{\bM\cdot\bM}{2B}\, ,\label{energy_balance_final}
\end{align}
and \eqref{force_balance_sleeve_edge_2} reduces to
\begin{align}
-\bS\cdot\bx'(0)
 = \bP\cdot\bx'(L)\, .\label{sleeve_edge_force_balance_final}
\end{align}
The relations \eqref{energy_balance_final} and \eqref{sleeve_edge_force_balance_final} are the results stated in equations (2.13) and (1.1) of \cite{Bigoni15}.  

We have chosen the solution that links smoothly with the fully static case.  Other solutions may exist if $\dot{s}_0$ vanishes.  If the rod is clamped instead of free to slide in a sleeve, the problem becomes that of a cantilever beam with the boundary conditions $\bR = -\bP$ and $\bM = -[\bx(L) - \bx(s_0)]\times\bP$.

\subsection{A peeling problem}\label{Sec:adhesion}

\begin{figure}[H]
	\centering
	\includegraphics[width=12cm]{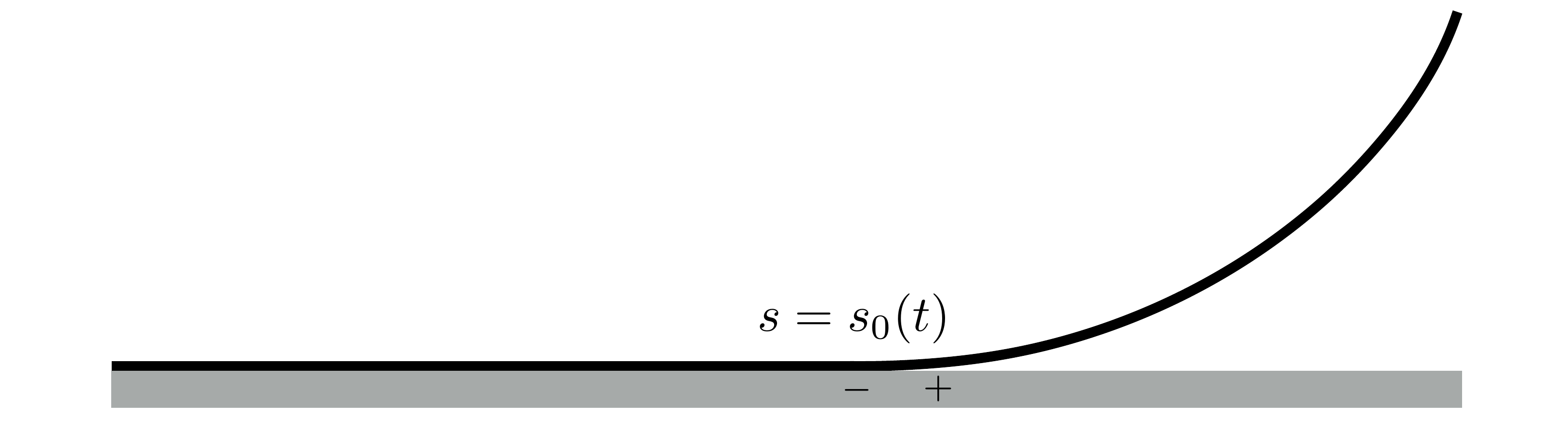}
	\captionsetup{margin=3cm}
	\caption{A rod $\bx(s,t)$ partially constrained by an adhesive surface.  Reactions at the contact discontinuity $s=s_0(t)$ may be present, but are not shown.}
	\label{Fig:elastica_peeling}
\end{figure}

Consider the arrangement in Figure \ref{Fig:elastica_peeling}, a much-studied problem  \cite{Bottega90, Kinloch1994, deBoerMichalske99, Plaut2001, Plaut2001-2, GlassmakerHui04, Majidi2007}.
A rod of uniform mass density is constrained by a flat, adhesive surface for $s\in[0,s_0]$.
Our interest is in the natural boundary condition for quasistatic peeling that relates bending and adhesion energies at the discontinuity.  

Unlike the sleeve problem, the peeling problem we study here entails irreversible processes, and entropy and dissipation can no longer be neglected. We first need an elementary model to justify a simple constitutive expression for $\eta$. One way of envisioning entropy in this problem is by imagining the adhesive as consisting of strong filamentous bonds with the substrate which can live in only two states: tight and loose. Tight bonds have no entropy, they cannot explore the space. Loose bonds have entropy, as they can take different shapes. Thus, the adhered rod has $\eta=0$, whereas the detached rod has a nonzero positive entropy $\eta>0$. Since no resistive forces hamper the detached portion of rod, the jump in entropy may only result from an entropy injection at the peeling singularity. Therefore, we make the following prescription for entropy,
\begin{align}
\eta = \eta_c\, ,
\end{align}
which, when combined with the constitutive prescription
\begin{align}
	\psi = \hat{\psi} + \psi_c = \tfrac{\bmo\cdot\bmo}{2B} + \psi_c \, , \quad\bmo=B\bOmega\, ,\label{peeling_prescriptions}
\end{align}
reduces equation \eqref{E_general_1} to
\begin{equation}
E = -\dot{s}_0\jump{\varepsilon_c}\,.\label{E_peeling_2}
\end{equation}
This relation describes $E$ as the rate of change of internal energy at the singularity. The quantity $\jump{\varepsilon_c}$ will soon be related to the adhesive energy.
If we still enforce $\tilde{E}=0$, the  balance of singular working \eqref{eq:A1} becomes
\begin{align}
W= \bR\cdot\add{\dot{\bx}} +\bM\cdot\add{\bomega}=\dot{s}_0\jump{\varepsilon_c} \, .\label{energy_balance_peeling_2}
\end{align}
The adhered region is stationary ($\dot{\bx}(s_0)=\zero$), with vanishing curvature ($\bOmega^-=\zero$), so the kinematics \eqref{linear_velocity_discontinuity} and  \eqref{angular_velocity_discontinuity} tell us that the velocity of the non-material peeling point $\bv_0 = \dot{s}_0\bx'(s_0)$ and the average material angular velocity $\add{\bomega} = -\dot{s}_0\tfrac{1}{2}\bOmega^+$.  
We now identify the term $\jump{\varepsilon_c} = 0 - (-\Gamma)$, where $\Gamma$ is a positive number representing the per length energy of adhesion between rod and surface. Thus, from \eqref{E_peeling_2}, we see that $E$ is a positive quantity, indicating just sufficient energy injection at the peeling front to break the adhesive bond.  Finally, using the singular moment balance \eqref{moment_jump} with $\bmo^-=0$, we transform \eqref{energy_balance_peeling_2} into
\begin{align}
\dot{s}_0\big(\Gamma - \tfrac{\bM\cdot\bM}{2B}\big)=0\, . 
\end{align}
Thus, for any quasi-static motion, 
\begin{align}
\Gamma=\tfrac{\bM\cdot\bM}{2B}\, .\label{peeling_boundary_condition}
\end{align}
The reaction moment vanishes in the absence of adhesion energy. Furthermore, the adhesion energy $\Gamma$ is subject to the constraint \eqref{dissipation_jump_3}, which may be written as \begin{align}
\Gamma\le\jump{\theta\eta_c}\, .
\end{align}
The relation \eqref{peeling_boundary_condition} is well known in the literature, where it is derived in various ways.  
Similar results can be derived for plates.  Majidi and Adams \cite{MajidiAdams09} showed that the effect of adhesion in partial contact of an elastic plate is equivalent to a discontinuity in the internal moment at the contact boundary. 
Hure and Audoly \cite{HureAudoly2013} derive the boundary condition at a two-dimensional elastocapillary peeling front; their equation (49g) is equivalent to our \eqref{peeling_boundary_condition}.
An informal approach equivalent to the present one applied to the peeling of a flexible adhesive tape may be found in Appendix C of a prior paper by two of the present authors \cite{SinghHanna2017}.

\section{Comparison with material force balance}\label{Sec:material_force}

In the preceding section, we derived results for two partial constraint problems for an \emph{elastica} using singular balances of force, moment, and energy \eqref{force_jump}-\eqref{energy_jump}, 
and some additional kinematic information.
In the present section, we discuss an alternate framework in which a singular balance of material force \eqref{material_force_jump} is used in lieu of the energy balance \eqref{energy_jump}.
O'Reilly proposed such a balance law for rods \cite{OReilly07}, compared it with the energy balance law \cite{OReilly2007}, and employed it in a number of settings including that of Bigoni's sleeve problems \cite{OReilly2015Eshelby, OReilly2017}.
What this approach must necessarily entail is a prescription for the source term $\config$ instead of the power $\tilde{E}$.

For the sleeve problem, $\config=0$, entropic terms are neglected, and the two conditions \eqref{energy_jump} and \eqref{material_force_jump} provide equivalent information.
Using the constitutive prescription \eqref{elastica_prescriptions}, the material jump \eqref{material_force_jump} can be written as
\begin{align}
\config + \jump{\bn\cdot\bx' + \tfrac{\bmo\cdot\bmo}{2B}} = 0\, .\label{material_force_jump_elastica}
\end{align}
The quantity inside the brackets is the conserved material force $c$.  The conservation of $c$ implies  $c^-= -\bS\cdot\bx'(0)$ and $c^+=\bP\cdot\bx(L)$, so that setting $\config=0$ recovers the result \eqref{sleeve_edge_force_balance_final}.    This appears to be the shortest possible derivation of this result. 
 Additionally, using the force and moment jumps \eqref{force_jump} and \eqref{moment_jump}, we can rewrite the material jump as
 \begin{align}
\config - \bR\cdot\add{\bx'} - \bM\cdot\add{\bOmega} = 0\, ,\label{material_force_jump_elastica_2}
\end{align}
and use constitutive relations and the kinematic information $\add{\bx'} = \bx'(s_0)$ and $\add{\bOmega} = \tfrac{1}{2}\bOmega^+$ to recover the other result \eqref{energy_balance_final} when $\config=0$.
Another manipulation is possible when $\bv_0=\zero$ and $\bomega_0=\zero$, as in the sleeve problem.  Using \eqref{linear_velocity_discontinuity} and \eqref{angular_velocity_discontinuity}, the energy jump in the form \eqref{driving_traction} can be rewritten $\tilde{E} +\dot{s}_0 \bR\cdot\add{\bx'} + \dot{s}_0\bM\cdot\add{\bOmega}=0$, and the identification $\tilde{E} = -\config\dot{s}_0$ made by comparison with \eqref{material_force_jump_elastica_2}.

However, our course of action is no longer clear when we consider the peeling problem in the presence of an adhesive energy. Majidi and co-workers  \cite{Majidi12,Majidi13, Majidi2007}  treated the peeling problem as purely mechanical, so that a distinction between free and internal energy did not arise, and accounted for the jump in adhesion energy in the source term $\config$. This assumption makes \eqref{material_force_jump_elastica} and \eqref{material_force_jump_elastica_2} valid for their treatment of the peeling problem as well.
They prescribe $\config = -\Gamma$, which when inserted into \eqref{material_force_jump_elastica_2} gives $\Gamma = \tfrac{\bM\cdot\bM}{2B} - \bR\cdot\bx'(s_0)$.  This differs from the correct result \eqref{peeling_boundary_condition} by an additional term $-\bR\cdot\bx'(s_0)$. Recognizing this, 
these authors assume the additional term to vanish.   
Note that in order to obtain the reactions in this problem, one needs to know the contact force  on the adhered side of the singularity, which requires some knowledge or assumptions about the behavior of the adhered portion of the rod. 
There is an alternate prescription for the material source term $\config$ that would deliver the right boundary condition, namely $\config = -\Gamma +\bR\cdot\bx'(s_0)$, but we see no physical argument that would lead us to this prescription other than our foreknowledge of what the answer should be.\footnote{Equation (34) of \cite{OReilly07} provides a relation between singular sources such that, given our prescription for $\tilde{E}$, the correct prescription for $Y$ can be obtained.  We thank O. M. O'Reilly for suggesting this approach.}
Using our energy balance approach, the phantom term simply does not arise.

\section{Concluding comments}\label{Sec:discussion}

The problems we have been discussing involve either the real or virtual motion of an internal boundary through a material.  
 The energy and material force balances are connected with symmetries in time and material coordinates, so there is a close relationship between them in problems of this type. 
  The role of kinematic jump conditions in an energy balance approach is taken by compatibility conditions on boundary variations in a ``configurational'' approach.
 
 These problems belong to the general class of free boundary problems \cite{BurridgeKeller1978}. 
 Unlike in the bulk region, the energy balance across such free boundaries is not redundant, but provides additional information about the system once some constitutive assumption regarding the net energy input at singularities has been made.  In the present cases, this assumption was the simple choice $\tilde{E}=0$, indicating in one case that the sleeve edge exerts a powerless system of reactions, and in the other that the peeling front injects exactly that energy which breaks the adhesive bond between rod and surface. 
 Alternatively, instead of the energy balance, the balance of material momentum and a constitutive assumption about material momentum injection can be used \cite{OReilly2017}.  
 
Both problems treated here illuminate our quasistatic theory for singular rods. Extension of the theory to the fully  dynamic case is called for. We anticipate that the dynamical counterpart of the reduced singular energy balance \eqref{driving_traction} will become an evolution law for the singularity.
Another interesting extension to the present topics involves smooth inhomogeneities in material properties in the bulk, as for example in recent work on variable-property rods in variable-curvature channels \cite{CicconofriDeSimone15, DalCorso17}.

\section*{Acknowledgments}
JAH and HS were supported by U.S. National Science Foundation grant CMMI-1462501. 	EGV acknowledges the kind hospitality of the Oxford Centre for
Nonlinear PDE, where part of this work was done while he was visiting the
Mathematical Institute at the University of Oxford.

\bibliographystyle{unsrt}

\end{document}